\documentclass[12pt,a4paper,keywords,prb]{revtex4-1}

\usepackage{graphicx}
\usepackage{amsmath}
\usepackage{epstopdf}
\usepackage{bm}
\usepackage[justification=raggedright]{caption}
\usepackage[labelfont=bf,textfont=normalfont,singlelinecheck=off,justification=raggedright]{subcaption}

\begin{document}

\title{Impurity bound states and disorder-induced orbital and magnetic order in the $s_\pm$ state of Fe-based superconductors}
\author{Maria N. Gastiasoro and Brian M. Andersen}
\affiliation{Niels Bohr Institute, University of Copenhagen, Universitetsparken 5, DK-2100 Copenhagen,
Denmark}

\begin{abstract}

We study the presence of impurity bound states within a five-band Hubbard model relevant to iron-based superconductors. In agreement with earlier studies, we find that in the absence of Coulomb correlations there exists a range of repulsive impurity potentials where in-gap states are generated. In the presence of weak correlations, these states are generally pushed to the edges of the gap, whereas for larger correlations the onsite impurity potential induces a local magnetic region which reintroduces the low-energy bound states into the gap. 

\end{abstract}

\date{\today}
\maketitle

\section{Introduction}

There are at least two reasons why the study of disorder effects in the high-T$_c$ superconductors remain an important topic. First, the superconducting state itself is generated by chemical doping which inevitably disorders the samples, and second, local probes of the quasi-particle states near the impurity sites can provide important information on the underlying system.\cite{balatsky06,alloul09} In the case of the cuprates, for example, it was shown how disorder acts to pin competing correlations, providing a natural explanation for the so-called spin-glass phase in the underdoped regime,\cite{andersen07,harter07,schmid10,andersen10} and STM measurements near isolated Ni impurities showed clear evidence for $d$-wave pairing symmetry of the superconducting order parameter.\cite{hudson01}  

In the iron-pnictides, several experimental scanning tunneling studies have been performed to investigate the modulations in the electronic spectrum caused by various defects.\cite{chuang10,song11,hoffman11,hanaguri12,allen12,yang12,grothe12} In the case of LiFeAs, for example, it has been recently shown how several kinds of defects exist on the surface, with distinct local structures in the local density of states (LDOS).\cite{grothe12} At present there is no theoretical model capturing the details of the LDOS near these different impurities. 

Theoretically, it is well-known that both potential and magnetic impurities can give rise to in-gap states in $d$-wave and multi-band $s_\pm$ superconductors.\cite{balatsky06,alloul09,andersen06} In the latter case, several theory studies of the single-impurity problem have been reported both with simplified two-band models,\cite{zhou09,bang09,tsai09,zhang09,li09,akbari10} and within a five-band approach.\cite{kariyado10} Here, we extend the study of the single-impurity problem within a five-band model including interactions treated in an unrestricted Hartree-Fock approximation. It is shown how an impurity locally induces orbital and magnetic order which can strongly modify the positions of the spectral in-gap bound states detectable by STM.   

\section{Model}

The five-orbital model Hamiltonian is given by
\begin{equation}
 \label{eq:H}
 H=H_{0}+H_{int}+H_{BCS}+H_{imp}.
\end{equation}
The first term is a tight-binding model,
\begin{equation}
 \label{eq:H0}
H_{0}=\sum_{\mathbf{ij},\mu\nu,\sigma}t_{\mathbf{ij}}^{\mu\nu}c_{\mathbf{i}\mu\sigma}^{\dagger}c_{\mathbf{j}\nu\sigma}-\mu_{0}\sum_{\mathbf{i}\mu\sigma}n_{\mathbf{i}\mu\sigma}.
\end{equation}
Here the operators $c_{\mathbf{i} \mu\sigma}^{\dagger}$ ($c_{\mathbf{i}\mu\sigma}$) create (annihilate) an electron at the $i$-th site in the orbital $\mu$ and with spin projection $\sigma$, and $\mu_{0}$ is the chemical potential.
The indices $\mu$ and $\nu$ run through 1 to 5 corresponding to the five $d_{xz}$, $d_{yz}$, $d_{x^2-y^2}$, $d_{xy}$ and $d_{3z^2}$ iron orbitals. 
The hopping integrals $t_{\mathbf{ij}}^{\mu\nu}$ are the same as those in Graser \emph{et al.}~\cite{graser09}, included up to fifth nearest neighbors.
Here, as elsewhere in this paper, the energy units are in electron volt (eV).  

The second term describes the onsite Coulomb interaction, 
\begin{align}
 \label{eq:Hint}
 H_{int}&=U\sum_{\mathbf{i},\mu}n_{\mathbf{i}\mu\uparrow}n_{\mathbf{i}\mu\downarrow}+(U'-\frac{J}{2})\sum_{\mathbf{i},\mu<\nu,\sigma\sigma'}n_{\mathbf{i}\mu\sigma}n_{\mathbf{i}\nu\sigma'}\\\nonumber
&\quad-2J\sum_{\mathbf{i},\mu<\nu}\vec{S}_{\mathbf{i}\mu}\cdot\vec{S}_{\mathbf{i}\nu}+J'\sum_{\mathbf{i},\mu<\nu,\sigma}c_{\mathbf{i}\mu\sigma}^{\dagger}c_{\mathbf{i}\mu\bar{\sigma}}^{\dagger}c_{\mathbf{i}\nu\bar{\sigma}}c_{\mathbf{i}\nu\sigma},
\end{align}
which includes the intra-orbital (inter-orbital) interaction $U$ ($U'$), the Hund's rule coupling $J$ and the pair hopping energy $J'$.
We will assume orbital and spin rotational invariance where the relations $U'=U-2J$ and $J'=J$ hold.

The third term is a phenomenological BCS pairing term
\begin{equation}
 H_{BCS}=-\sum_{\mathbf{i}\neq \mathbf{j},\mu\nu}[\Delta_{\mathbf{ij}}^{\mu\nu}c_{\mathbf{i}\mu\uparrow}^{\dagger}c_{\mathbf{j}\nu\downarrow}^{\dagger}+H.c.],
\end{equation}
with the superconducting (SC) order parameter $\Delta_{\mathbf{ij}}^{\mu\nu}=V_{\mathbf{ij}}\langle c_{\mathbf{j}\nu\downarrow}c_{\mathbf{i}\mu\uparrow}\rangle$ where $V_{\mathbf{ij}}$ denotes the strength of the effective attraction. 
The pairing is chosen as next-nearest-neighbor intra-orbital pairing, which reproduces the fully gapped $s_{\pm}$ state.

The last term in the Hamiltonian is a nonmagnetic impurity term
\begin{equation}
 H_{imp}=V_{imp}\sum_{\mathbf{i^*}\mu\sigma}c_{\mathbf{i^*}\mu\sigma}^{\dagger}c_{\mathbf{i^*}\mu\sigma},
\end{equation} 
which adds a local potential $V_{imp}$ at a single site $\mathbf{i^*}$ in all five orbitals, neglecting the orbital dependence for simplicity.

After a mean-field decoupling of the onsite interaction term \eqref{eq:Hint} in both the ``density'' and the ``Cooper'' channels, a Bogoliubov transformation results in the following multi-orbital Bogoliubov de-Gennes equations

\begin{align}
\sum_{\mathbf{j}\nu}
\begin{pmatrix}
H_{\mathbf{i}\mu \mathbf{j}\nu \sigma} & \Delta_{\mathbf{i}\mu \mathbf{j}\nu}\\
\Delta_{\mathbf{i}\mu \mathbf{j}\nu}^{*} & -H_{\mathbf{i}\mu \mathbf{j}\nu \bar{\sigma}}^{*} 
\end{pmatrix}
\begin{pmatrix}
 u_{\mathbf{j}\nu}^{n} \\ v_{\mathbf{j}\nu}^{n} 
\end{pmatrix}=E_{n}
\begin{pmatrix}
 u_{\mathbf{i}\mu}^{n} \\ v_{\mathbf{i}\mu}^{n} 
\end{pmatrix},
\end{align}
where
\begin{align}
 H_{\mathbf{i}\mu \mathbf{j}\nu \sigma}&=t_{\mathbf{ij}}^{\mu\nu}+\delta_{\mathbf{ij}}\delta_{\mu\nu}[-\mu_{0}+\delta_{\mathbf{ii^*}}V_{imp}+U \langle n_{\mathbf{i}\mu\bar{\sigma}}\rangle\\\nonumber
&\quad+\sum_{\mu' \neq \mu}(U'\langle n_{\mathbf{i}\mu' \bar{\sigma}}\rangle+(U'-J)\langle n_{\mathbf{i}\mu' \sigma}\rangle)],
\end{align}
and
\begin{align}
 \Delta_{\mathbf{i}\mu \mathbf{j}\nu}&=\delta_{\mathbf{ij}}\delta_{\mu\nu}[\Delta_{\mathbf{ii}}^{\mu\mu(U)}+2\sum_{\mu' \neq \mu}\Delta_{\mathbf{ii}}^{\mu'\mu' (J')}]\\\nonumber
&\quad+2\delta_{\mathbf{ij}}\sum_{\mu' \neq \mu}[\Delta_{\mathbf{ii}}^{\mu\mu'(U')}+\Delta_{\mathbf{ii}}^{\mu' \mu(J)}]-\Delta_{\mathbf{ij}}^{\mu\nu}.
\end{align}
The local densities and the SC order parameters are obtained through the following self-consistency equations 
\begin{align}
  \langle n_{\mathbf{i}\mu\uparrow} \rangle&=\sum_{n}|u_{\mathbf{i}\mu}^{n}|^{2}f(E_{n}),\\\nonumber
  \langle n_{\mathbf{i}\mu\downarrow} \rangle&=\sum_{n}|v_{\mathbf{i}\mu}^{n}|^{2}(1-f(E_{n})),\\
  \Delta_{\mathbf{ii}}^{\mu\nu(X)}&=X\sum_{n}u_{\mathbf{i}\mu}^{n}v_{\mathbf{i}\nu}^{n*}f(E_{n}),\\
  \Delta_{\mathbf{ij}}^{\mu\nu}&=V_{ij}\sum_{n}u_{\mathbf{i}\mu}^{n}v_{\mathbf{j}\nu}^{n*}f(E_{n}),
\end{align}
where $X=U$, $U'$, $J$ or $J'$.

\section{Results and Discussion}

We focus on signatures associated with the $s_\pm$ state, generated by an intra-orbital superconducting pairing $V_{\bf ij}=0.65$ between next-nearest neighbor sites. 
The chemical potential $\mu_0$ is fixed so that the total density is $n=6.1$ (electron-doped), where the system is in the SC state but close to the SDW phase, and a non-magnetic impurity is placed at $(x_{imp},y_{imp})=(14,14)$ in a $28\times28$ lattice. 

\begin{figure}[b]
\begin{subfigure}{0.32\textwidth}
 \subcaption{}
 \includegraphics[clip=true,width=0.99\columnwidth]{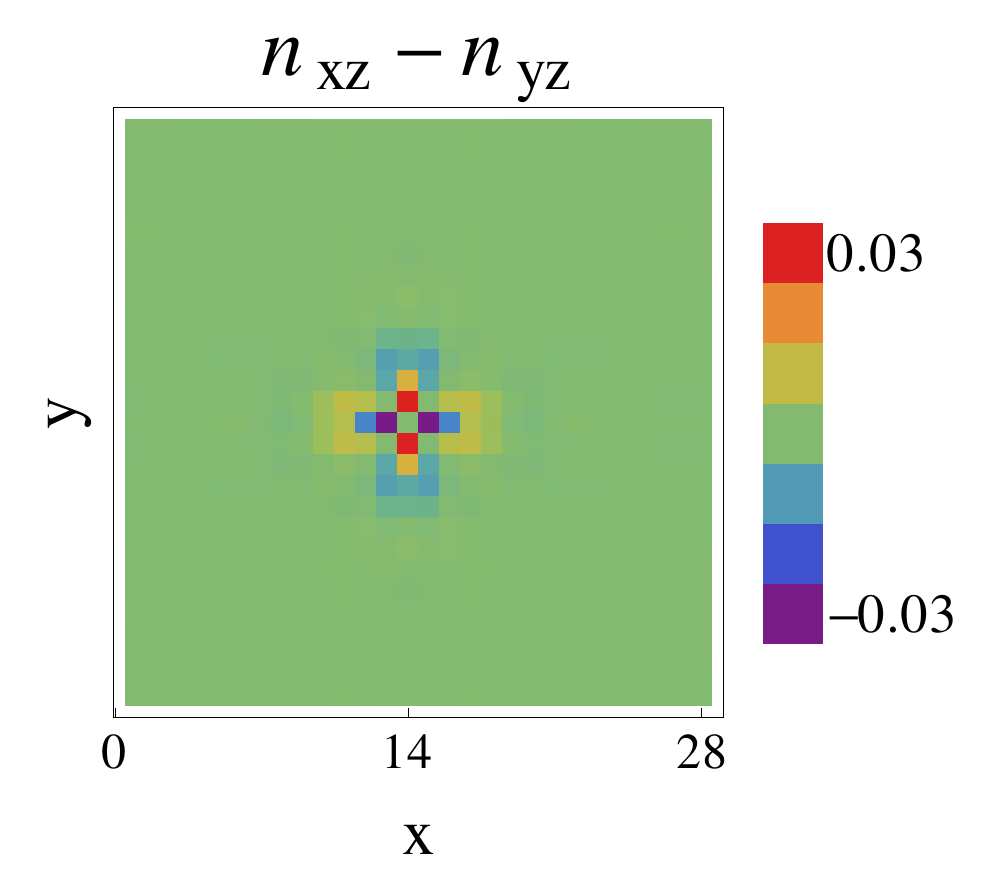}
\end{subfigure}
\begin{subfigure}{0.32\textwidth}
 \subcaption{}
 \includegraphics[clip=true,width=0.99\columnwidth]{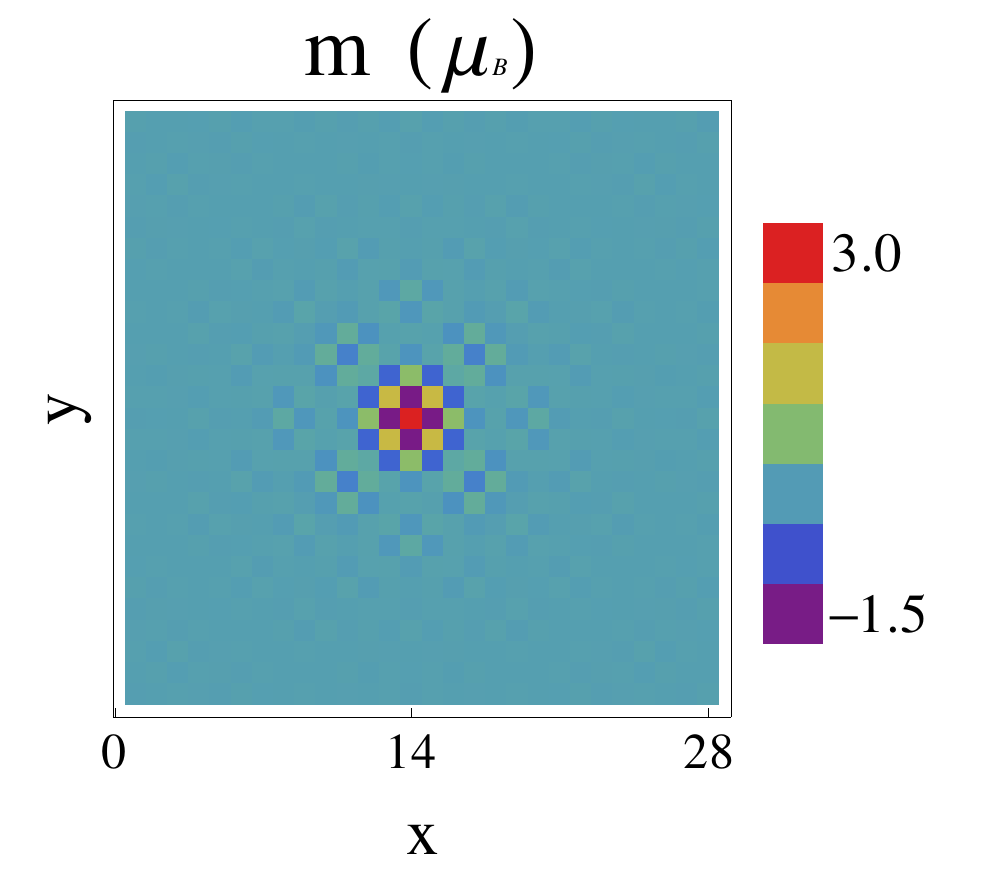}
\end{subfigure}
\begin{subfigure}{0.32\textwidth}
 \subcaption{}
 \includegraphics[clip=true,width=0.99\columnwidth]{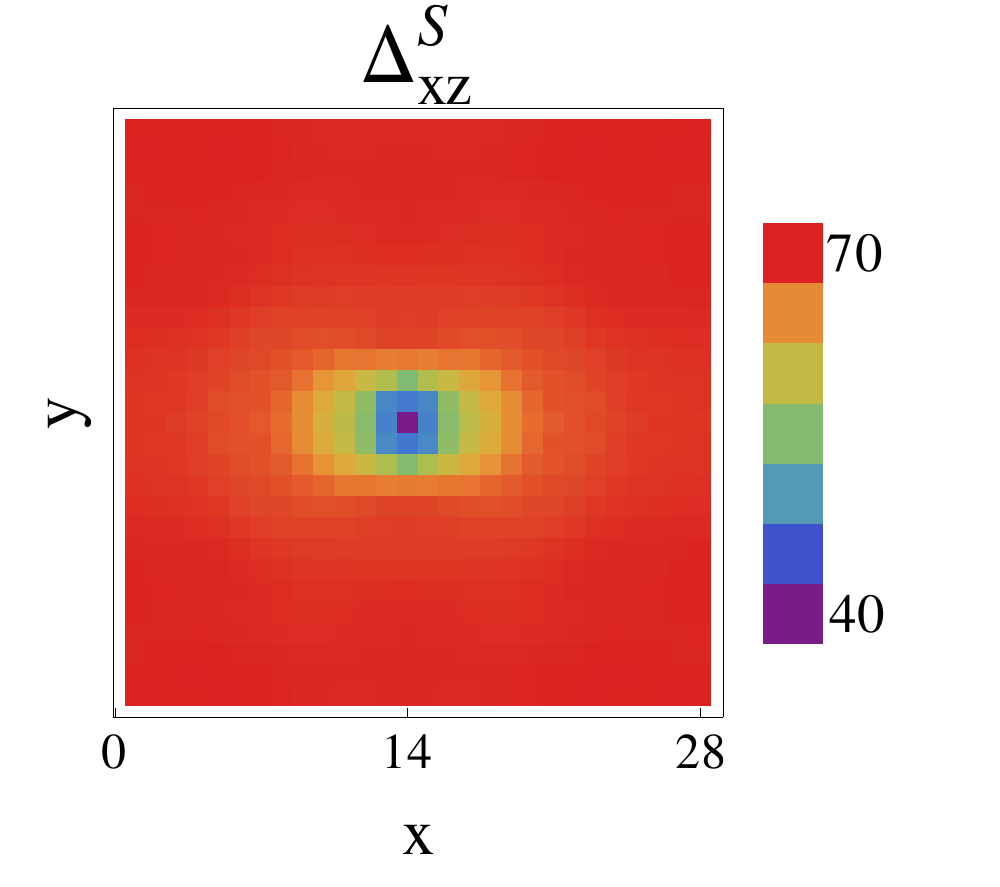}
\end{subfigure}
\caption{Real-space distribution of the self-consistent mean fields for a repulsive impurity ($V_{imp}=1$). 
The interacting parameters have been chosen as $U=1.35$ and $J=U/4$. 
Induced local (a) orbital ordering, (b) magnetization and (c) suppression of the superconducting singlet component of the $d_{xz}$ orbital in meV.} 
\label{fig:oo-m-dx}
\end{figure}

For repulsive potentials ($V_{imp}>0$), there is an energy penalty for electrons to jump on to the impurity site.
From the orientations of the orbitals $d_{xz}$ and $d_{yz}$ at each Fe site, their hopping amplitudes $t_{\bf ij}^{\mu\nu}$ along x and y directions are the same but rotated by $\pi/2$ with respect to each other. 
Therefore, the effective ``forbidden hopping'' effect of the impurity is reflected in $d_{yz}$ with a $\pi/2$ rotation with respect to the same effect in $d_{xz}$, and a local twofold orbital ordering is induced around the impurity. 
Figure \ref{fig:oo-m-dx}(a) shows the orbital ordering around a repulsive impurity.
For attractive potentials ($V_{imp}<0$), a similar orbital ordering is induced, simply rotated by $\pi/2$.  

Above a critical value $U_c$ (with $J=U/4$ fixed), local magnetization is also induced around repulsive impurities. 
An example of the real space distribution of the magnetic order parameter $m_{\bf i}=\sum_\mu (n_{{\bf i}\mu\uparrow}-n_{{\bf i}\mu\downarrow})$ $\mu_B$ is plotted in figure \ref{fig:oo-m-dx}(b).
For attractive potentials on the contrary, no induced magnetization is found. 

\begin{figure}[b]
  \begin{subfigure}{0.49\textwidth}
   \subcaption{}
     \includegraphics[width=0.65\textwidth]{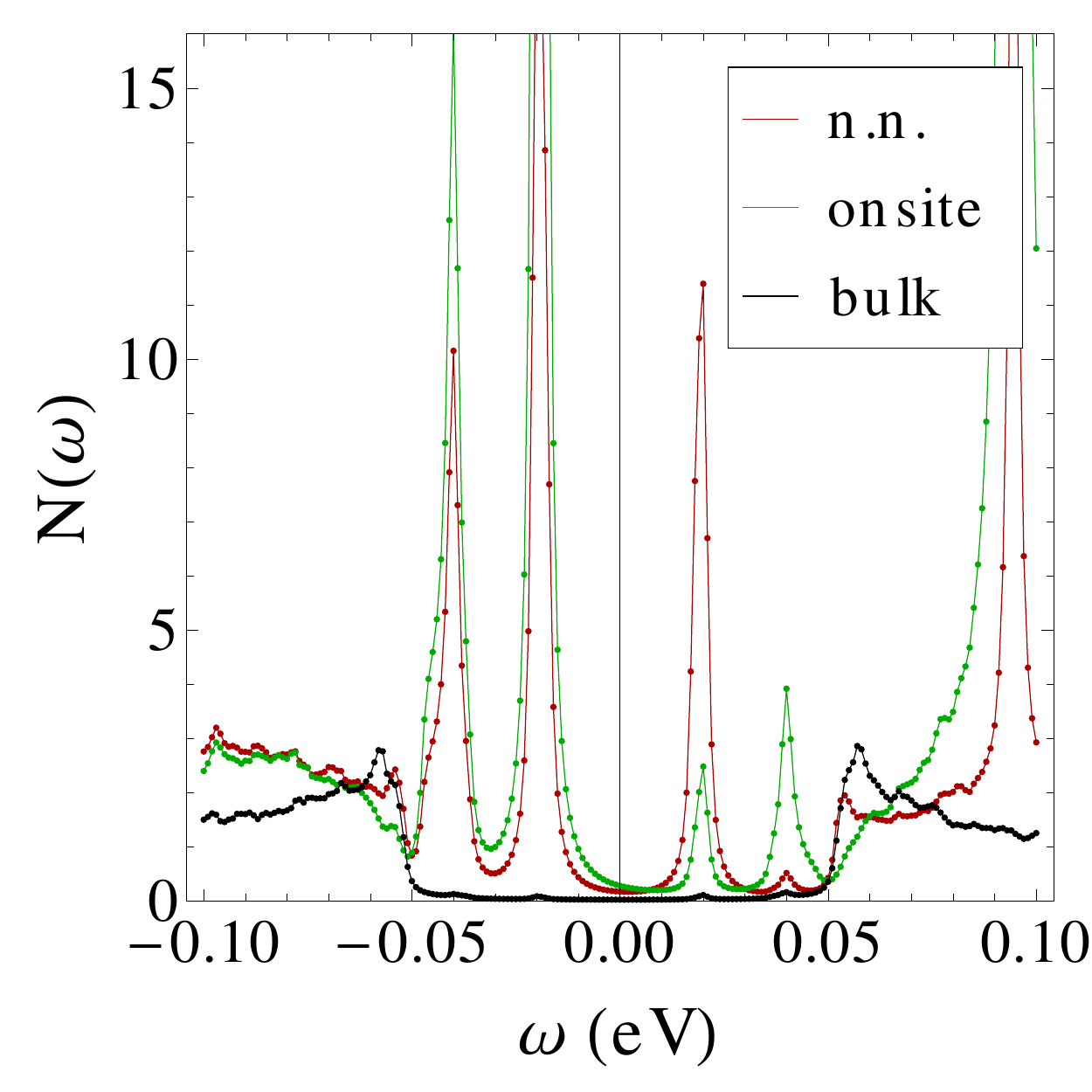}
  \end{subfigure}
  \begin{subfigure}{0.49\textwidth}
   \subcaption{}
     \includegraphics[width=0.65\textwidth]{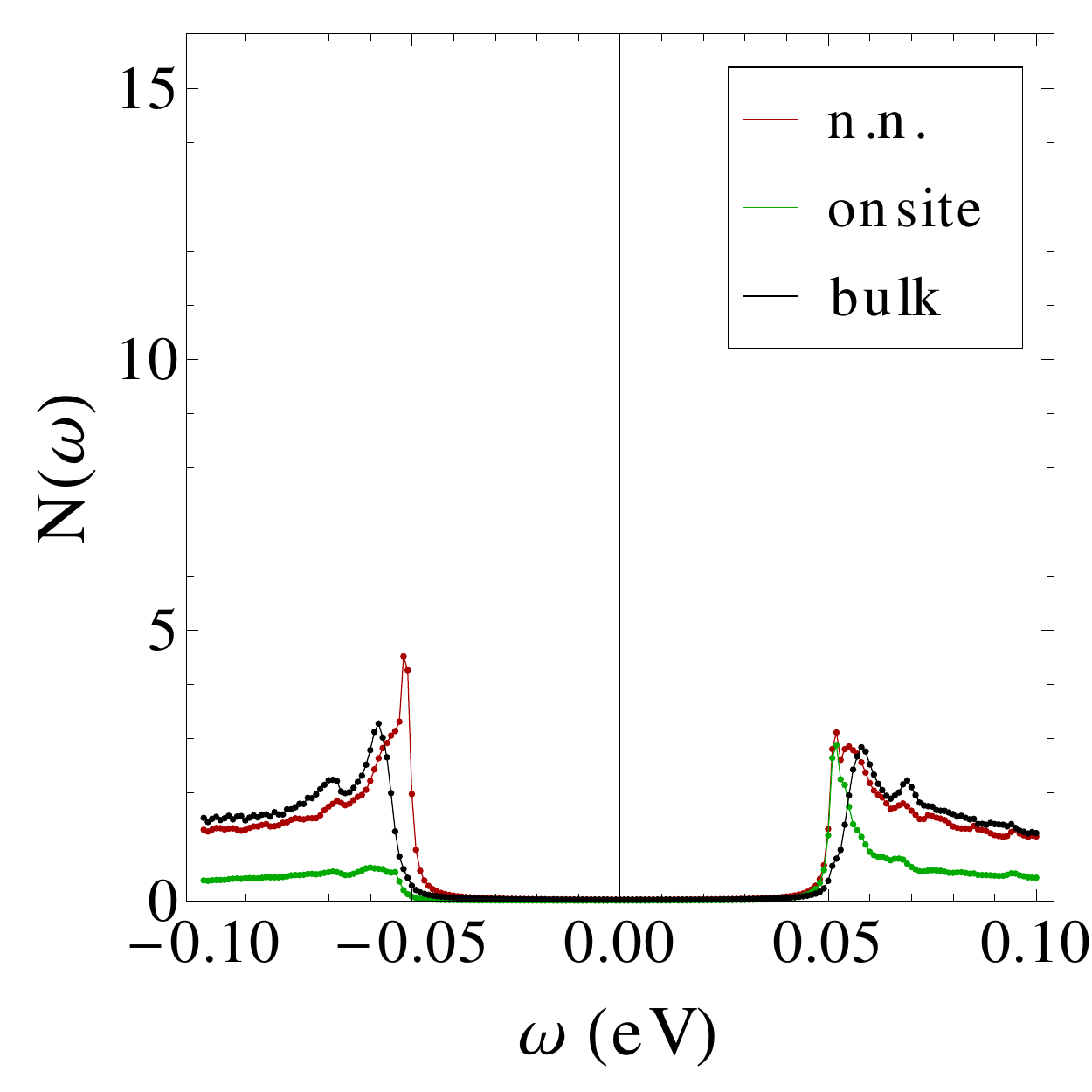}
  \end{subfigure}
 \caption{LDOS for a (a) repulsive ($V_{imp}=1$) and (b) attractive ($V_{imp}=-1$) impurity. 
The interaction parameters are chosen to be $U=J=0$. 
The black curve is the DOS far away from the impurity site and the red and green curves the DOS at the nearest neighbor and impurity site, respectively.}
\label{fig:ldos-bs}
\end{figure}

Let us analyze why this effect depends on the type of impurity. 
The local density of states (LDOS) at site ${\bf i}$ is given by
 \begin{equation}
 N_{\bf i}(\omega)=-\frac{1}{\pi}Im\sum_{n\mu}(\frac{|u_{{\bf i}\mu}^{n}|^{2}}{\omega-E_{n}+i\eta}+\frac{|v_{{\bf i}\mu}^{n}|^{2}}{\omega+E_{n}+i\eta}),
\end{equation}
and calculated using the ``supercell'' method, with a $25 \times 25$ copies of the original $28 \times 28$ lattice, in order to obtain high spectral resolution with $\eta=0.001$.
Figure \ref{fig:ldos-bs} shows the LDOS for both types of impurities in the $U=J=0$ uncorrelated case.
Around the repulsive impurity, states are generated inside the SC gap; on the contrary, the gap is almost unchanged and clean around the attractive impurity. 
These results agree with a previous five-orbital single-impurity study.\cite{kariyado10} 
In the static long wavelength limit ($\omega=0$, $k\rightarrow0$), the real part of the bare spin susceptibility $\chi_0(k\rightarrow0,0)$ is proportional to the DOS at the Fermi level, and the Stoner criterion becomes $U N(E_F)\rightarrow1$.
Because of the presence of bound states in the case of repulsive impurities, the DOS is generally higher at the Fermi level, allowing the Stoner instability to be crossed locally around the impurity site.\cite{astrid}

Figure \ref{fig:oo-m-dx}(c) shows the obtained SC order parameter and its modulation around the impurity site.
A representative singlet component for each orbital $\mu$ is given by, 
\begin{equation}
 \Delta_{{\bf i}\mu}^S=\frac{1}{2}\sum_{\bf j} (\Delta_{\bf ij}^{\mu\mu}-\Delta_{\bf ji}^{\mu\mu}).
\end{equation}
The spatial modulation of this orbitally resolved SC order parameter around the impurity follow the symmetry of their corresponding orbitals; twofold symmetry for $\Delta_{xz}^S$ and $\Delta_{yz}^S$, and fourfold for the rest of the orbitals.

\begin{figure}[b]
 \centering
  \begin{subfigure}{0.3\textwidth}
   \subcaption{}
     \includegraphics[width=0.99\textwidth]{ldos_U0.pdf}
  \end{subfigure}
  \begin{subfigure}{0.3\textwidth}
   \subcaption{}
     \includegraphics[width=0.99\textwidth]{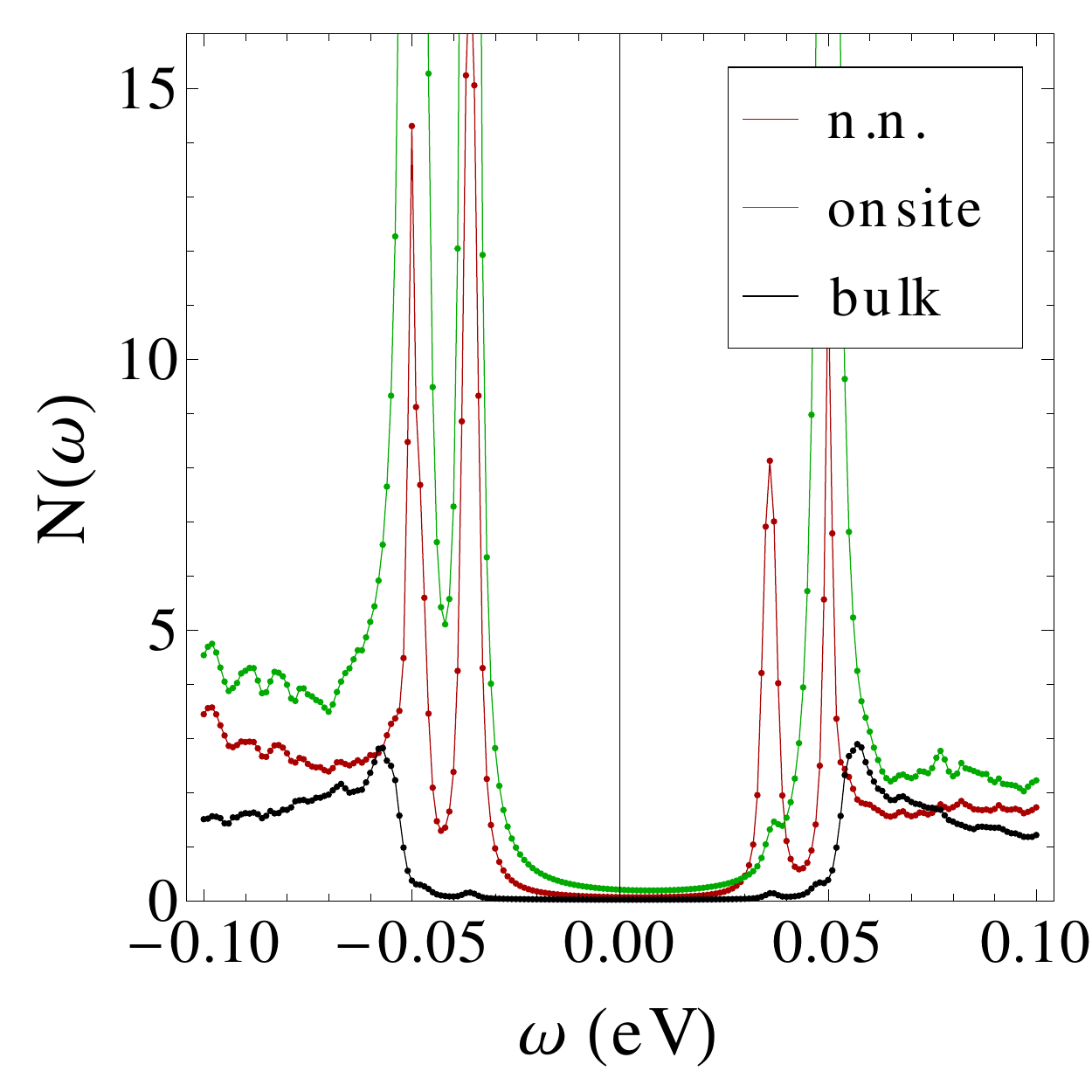}
  \end{subfigure}
  \begin{subfigure}{0.3\textwidth}
   \subcaption{}
     \includegraphics[width=0.99\textwidth]{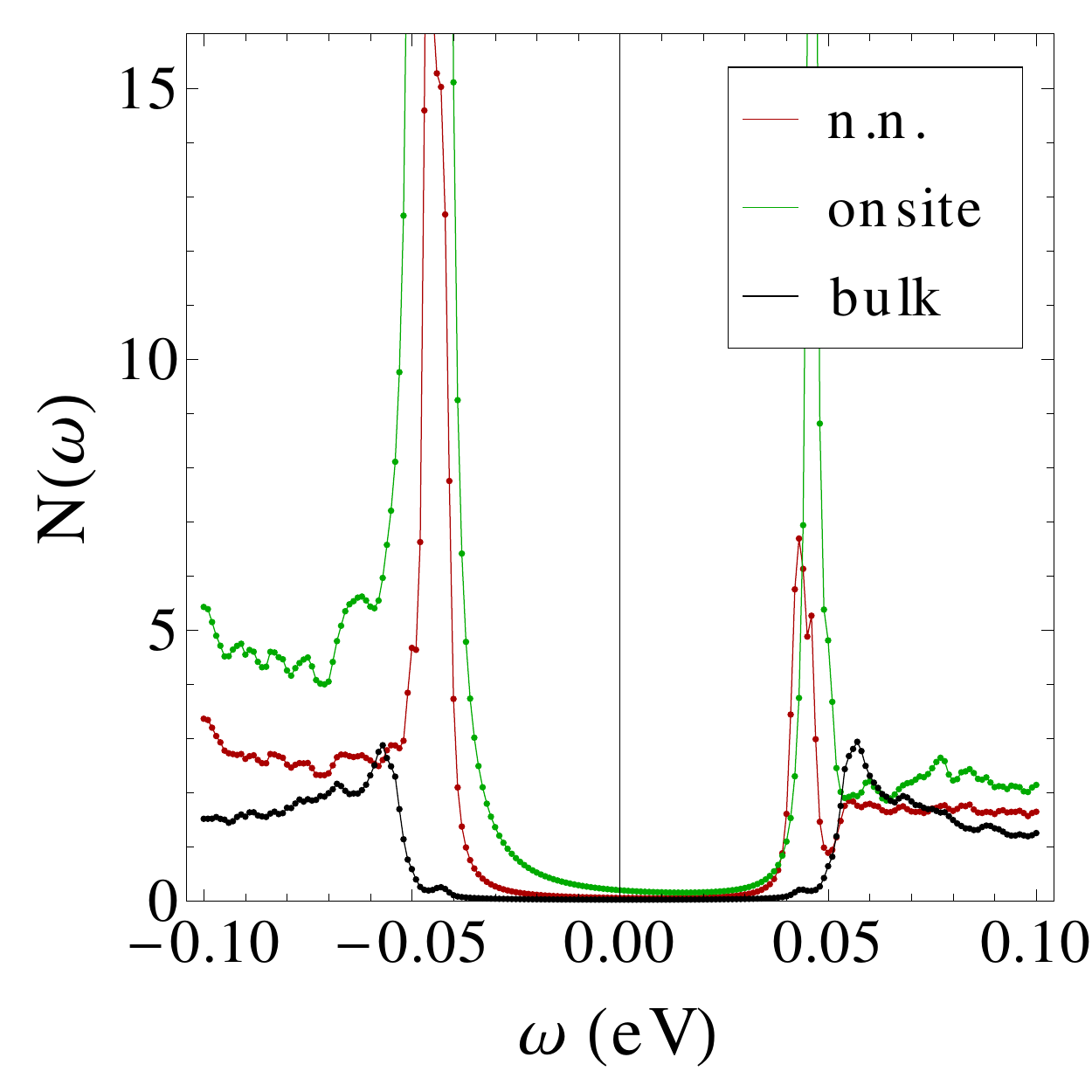}
  \end{subfigure}
  \begin{subfigure}{0.3\textwidth}
   \subcaption{}
     \includegraphics[width=0.99\textwidth]{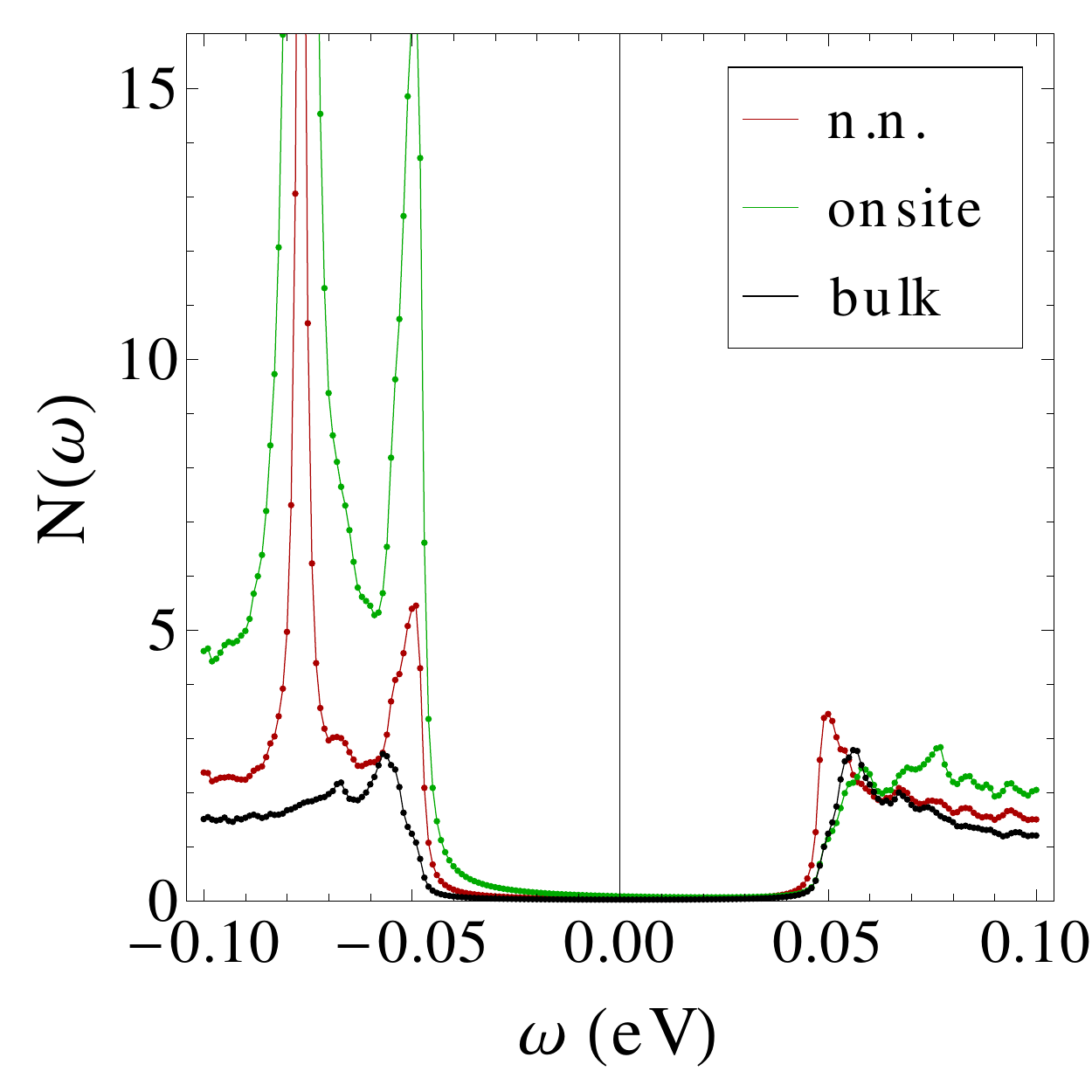}
  \end{subfigure}
  \begin{subfigure}{0.3\textwidth}
   \subcaption{}
     \includegraphics[width=0.99\textwidth]{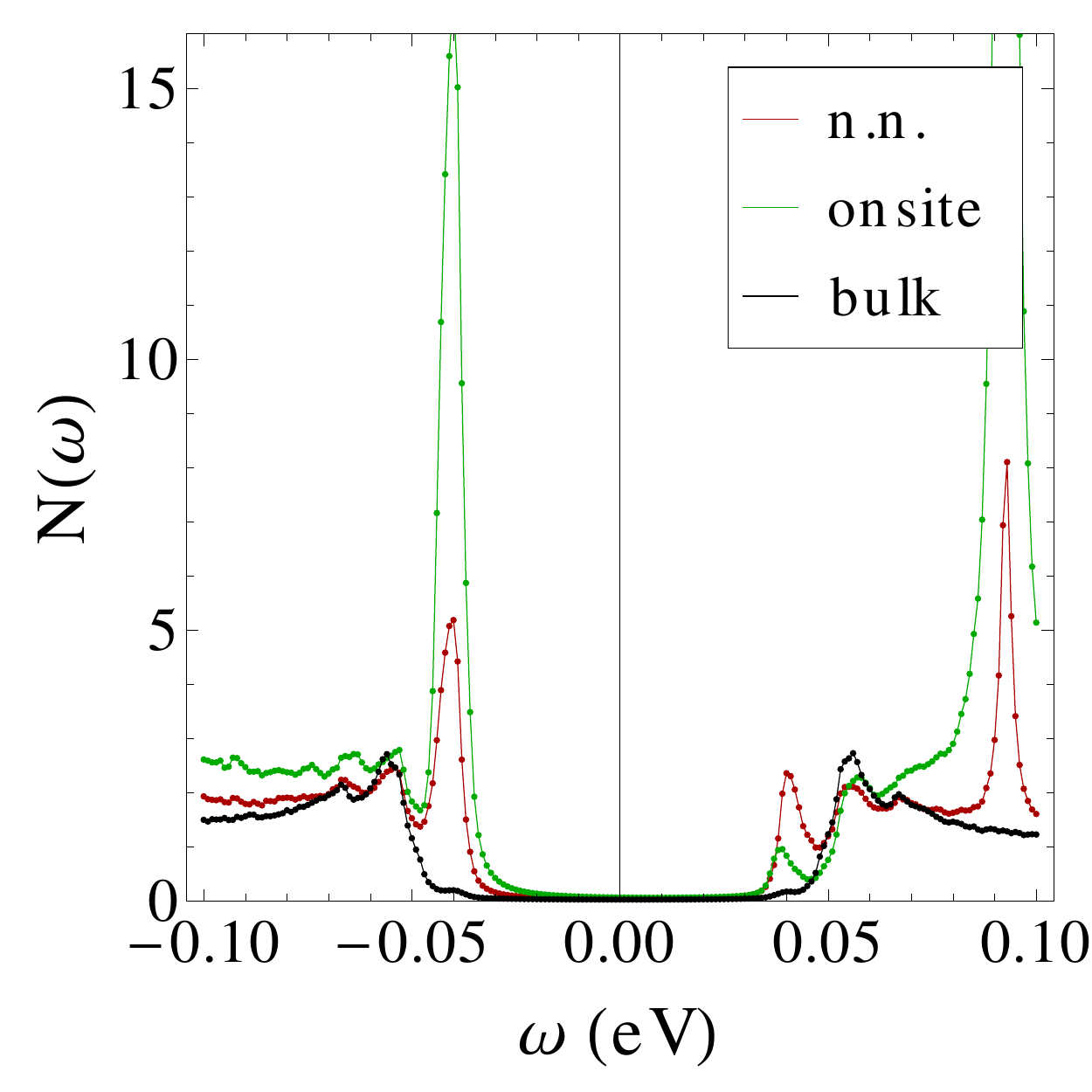}
  \end{subfigure}
  \begin{subfigure}{0.3\textwidth}
   \subcaption{}
     \includegraphics[width=0.99\textwidth]{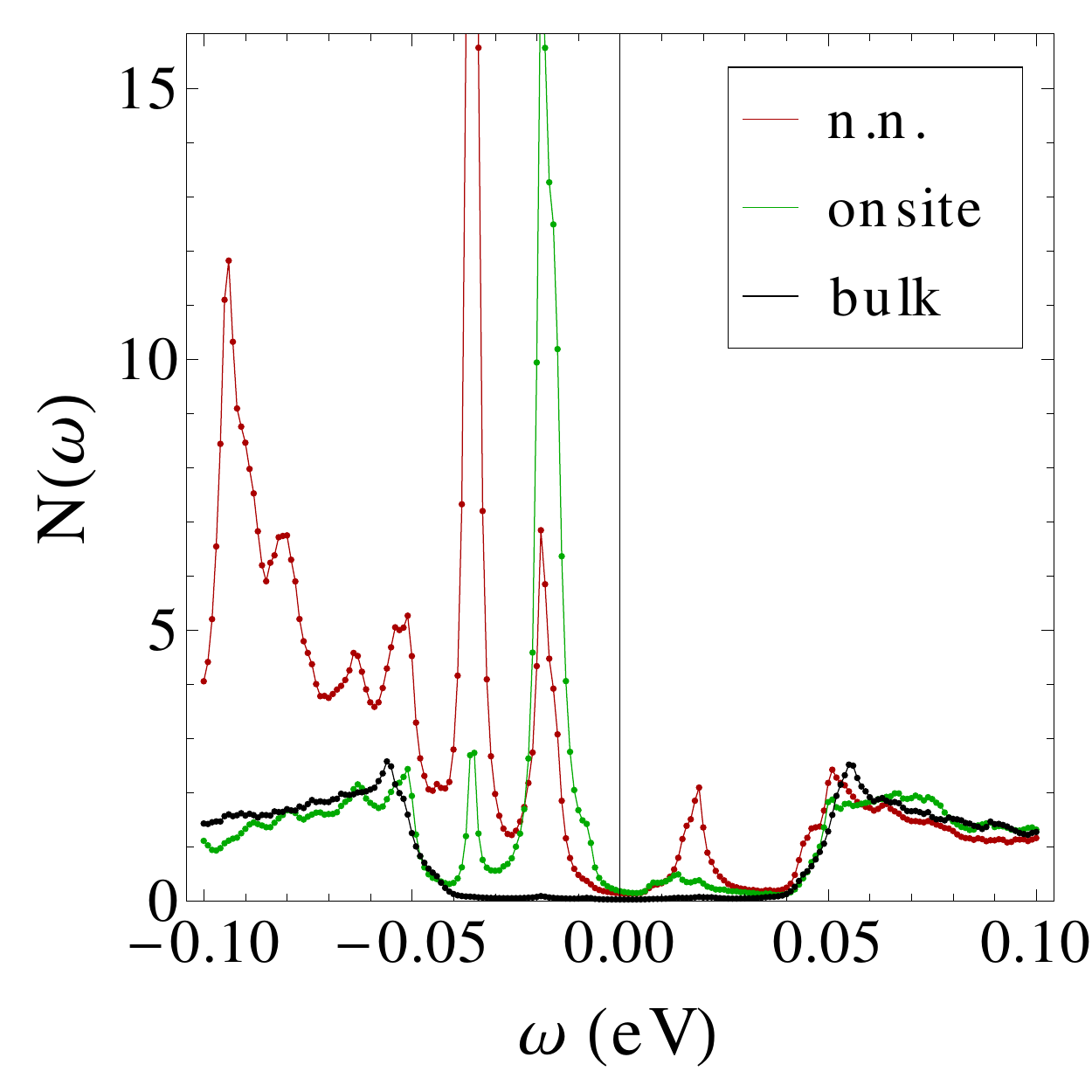}
  \end{subfigure}
\caption{LDOS around a repulsive impurity ($V_{imp}=1$), for various correlation strengths $U$ and $J=U/4$.
(a) $U=0$, (b) $U=0.2$, (c) $U=0.4$, (d) $U=1.0\lesssim U_c=1.1$, (e) $U=1.2\gtrsim U_c$ and (f) $U=1.35>U_c$. The black curve is the DOS far away from the impurity site and the red and green curves the DOS at the nearest neighbor and impurity site, respectively.}
\label{fig:ldos}
\end{figure}

Finally, we analyze the role of the strength of the correlations on the final LDOS.
We focus on repulsive potentials, where local in-gap bound states are generated and magnetization can be induced around the impurity.
The correlation strength dependence of the LDOS is summarized in figure \ref{fig:ldos}. 
The uncorrelated $U=J=0$ case can be seen in figure \ref{fig:ldos}(a). 
There are four in-gap bound states at the nearest-neighbor and impurity sites.
When the correlations are slightly increased, these states are pushed away from the Fermi level (panels \ref{fig:ldos}(b)-(c)). 
At values of $U$ below the critical Stoner $U_c$, the impurity-state formation happens at the edges or outside of the gap.
For higher strength of correlations, $U$ starts getting close to the critical value ($U\rightarrow U_c$), and in-gap states are formed again. 
It is then possible for the system to locally cross the Stoner instability, and magnetization is induced.
Finally, the high correlation case $U>U_c$, is shown in figure \ref{fig:ldos}(f). 
Strong magnetization sets in in the vicinity of the potential (see figure \ref{fig:oo-m-dx}(b)), and new local in-gap magnetic features appear in the LDOS around the Fermi level associated with the effective cluster of "magnetic" impurities surrounding the non-magnetic potential.

\section{Conclusions}

In summary, we have studied the single-impurity problem in an effective five-orbital Hubbard model with
interactions included at the mean field level. Superconductivity is included by a phenomenological BCS term 
where pairing between next-nearest neighbors generate a fully gapped s$_\pm$ state.  

Local properties such as orbital ordering and magnetization are induced around the impurity potentials.
The orbital ordering appears around repulsive and attractive potentials, because of an effective hopping asymmetry of the ordered orbitals.
Magnetization is induced around repulsive scatterers. 
These kind of impurities are pair-breaking and develop local in-gap bound states, enhancing the LDOS around the Fermi level. 
The Stoner condition can then be locally satisfied for strong enough correlations.  
By contrast, attractive impurities have an almost uniform clean gap, and do not induce local magnetization. 

Finally, we discuss the role of correlations on the impurity bound states. At low correlation strengths, the bound states tend to be pushed out of the SC gap. 
However, when the correlation strength approach a critical value $U_c$, the in-gap bound states are pushed back into the gap, and finally when local magnetization is induced around the potential additional sub-gap peaks appear in the LDOS resulting from the magnetization.

\section{Acknowledgement}

B.M.A. acknowledges support from The Danish Council for Independent Research $|$ Natural Sciences.

\end{document}